     \documentstyle[12pt]{article}   
     \newlength{\dinwidth}                       
     \newlength{\dinmargin}                      
\def\lsim{\mathrel{\rlap{\lower4pt\hbox{\hskip1pt$\sim$}}
    \raise1pt\hbox{$<$}}}                
\def\gsim{\mathrel{\rlap{\lower4pt\hbox{\hskip1pt$\sim$}}
    \raise1pt\hbox{$>$}}}                
\def\Pom{{\bf I\!P}}
\begin{document}
\vspace*{10mm}
\begin{center}  \begin{Large} \begin{bf}Intrinsic $k_{\perp}$
  in the pomeron\\
  \end{bf}  \end{Large}
  \vspace*{5mm}
  \begin{large}
N.N. Nikolaev\\
  \end{large}
IKP(Theorie), FZ J{\"u}lich, D-52428 J{\"u}lich, Germany \\
\&
L. D. Landau Institute for Theoretical Physics, GSP-1,
117940, \\
ul. Kosygina 2, Moscow 117334, Russia\\
\end{center}
\begin{quotation}
\noindent
{\bf Abstract:}
In a consistent QCD approach to diffractive DIS gluons in the pomeron
have soft $z_{g}$ distribution and very broad intrinsic $k_{\perp}$ 
distribution. The latter has not yet been incorporated into diffractive
Monte-Carlo generators (RAPGAP,....). I argue that the so-introduced bias
may be the reason why H1 Collaboration finds that the diffractive jet 
analysis calls for very large and hard gluon 
content of the pomeron. I discuss
possible direct experimental determinations of intrinsic $k_{\perp}$ in
the pomeron.
\end{quotation}

\noindent
{\bf 1.~~ Glue and charged partons in the pomeron: the QCD viewpoint}\\

The DGLAP evolution analysis to diffractive DIS structure function 
$F_{2}^{D(3)}$ has led the H1 Collaboration to a dramatic conclusion 
\cite{H1} of {\bf very strong and hard glue} in the pomeron.  

We recall that there are several reasons why conclusions derived from 
the analysis \cite{H1} are questionable. First, it is well established
that the Regge factorization used in \cite{H1} does not hold 
\cite{GNZ95,GNZcharm}. Second, there is a fundamental difference of
cancelations of virtual and real radiative corrections in
inclusive vs. diffractive DIS \cite{NZ94}, the analysis of which 
to our opinion has not been exhausted in the existing discussions
of hard factorization for diffraction \cite{Collins} thus leaving 
the status of DGLAP evolution for diffractive DIS open. At $\beta 
\sim 1$ the DGLAP evolution has never been proved, it has rather been 
questioned  \cite{GNZcharm} because the intrinsic size of the pomeron 
varies with $\beta$. Third, a proof of the DGLAP evolution 
to diffractive structure $F_{2}^{D(3)}$ exists \cite{NZ94}
only for $\beta \ll 1$
and only to leading $\log Q^{2}$.
Fourth, even for $\beta \ll 1$ it only holds for fixed $x_{\Pom}$ 
because the intrinsic structure of the pomeron varies with $x_{\Pom}$
\cite{GNZ95,GNZcharm,NZ94} which is still another manifestation of the lack of 
Regge factorization.  Fifth, the higher twist effects in $F_{2}^{D(3)}$ 
are abnormally large especially at large $\beta$ \cite{GNZlong,Twist4}. 
To this end we recall that
the conclusions of the analysis \cite{H1} on the glue in pomerons
are especially sensitive to $F_{2}^{D(3)}$ at large $\beta$.

All these reservations notwithstanding, one can still address the issue 
of the quark-gluon content of the pomeron
on a consistent QCD basis in which diffractive DIS
is described by excitation of multiparton Fock states of the photon
\cite{NZ94,NZ92}. The lowest order QCD process - excitation of the 
quark-antiquark state of 
the photon -  can be reinterpreted as DIS off the quark-antiquark valence
state of the pomeron. The next-to-lowest order QCD process - excitation 
of the quark-antiquark-gluon state of the photon - can be reinterpreted 
as DIS off the quark-antiquark sea which is generated from the two-gluon
valence state of the pomeron, i.e., the photon-gluon fusion.
 Excitation of still higher order Fock
states of the photon can be reinterpreted - with certain 
reservations and within certain limitations - as QCD evolution of
diffractive structure function built upon those valence parton distributions. 
In such a consistent QCD theory of diffractive DIS the quark-glue content
of the pomeron depends on $x_{\Pom}$, but at the values of $x_{\Pom}$ of
the practical interest one unequivocally predicts a glue which is soft
and carries about the same momentum as charged partons \cite{GNZ95,NZ94,NZ92}
in variance with the H1 conclusions. 

Consequently, the real issue is an experimental separation of the
quark-antiquark and quark-antiquark-gluon final states. Such a 
separation of  photon-gluon fusion has been initiated by H1 collaboration
which found a consistency of the observed jet activity with their hard 
and large glue in the pomeron. However, such an analysis of the jet activity 
is not a model-independent one.
Here I make a point that the version of the 
RAPGAP MC code the H1 analysis has been based upon 
does not include the intrinsic transverse momentum of partons in the
pomeron ({\sl H. Jung, private communication}) which may be a reason 
why within the present version of RAPGAP the jet
analysis tends to ask for a large and hard density of gluons in
the pomeron, much larger and harder than predicted by QCD models of
diffractive DIS \cite{GNZ95,NZ94}.\\

\noindent
{\bf 2.~~ Standard collinear parton model misses 
dynamical intrinsic $k_{\perp}$ of partons}\\

The central point of the standard $Q^{2}$-factorization for `hard' 
processes is that the underlying `hard' partonic QCD cross section 
can be computed assuming that the incoming partons have {\bf negligible} 
transverse momenta compared to the relevant hard scale be it $Q$ or jet 
transverse momentum $p_{J}$, which for reasons explained below is  
a poor approximation for jet 
production. For instance, the $Q^{2}$-factorization jet cross
section for photon-gluon fusion subprocess is proportional to the gluon 
structure function 
$G(\bar{x},\sim p_J^{2})$  which according to Fadin, Kuraev and 
Lipatov (FKL  \cite{FKL}) is an integral flux {\bf of all gluons in the proton 
with the transverse momenta $0 < k_{\perp}  \lsim p_{J}$},
\begin{equation}
G(\bar{x}, p_J^{2})= \int_{0}^{p_{J}^{2}} {d k_{\perp}^{2}\over k_{\perp}^{2}} 
{dG(\bar{x}, k_{\perp}^{2}) \over d\log k_{\perp}^{2} }\, .
\label{eq:1}
\end{equation} 
This intrinsic $k_{\perp}$ of partons originates 
from dynamical QCD evolution. The experimentally observed 
strong scaling violations tell that 
 $\left. dG(\bar{x},k_{\perp}^{2})/d\log k_{\perp}^{2}\right|_{k_{\perp}=p_{J}}$ 
is substantial and it would be illegitimate to neglect 
$k_{\perp} \sim p_{J}$ in the consistent calculation of the jet cross section,
\begin{equation}
d\sigma_{J}(\vec{p}_{J})\sim \int{ dx_{g}\over x_{g}}
\int {d^{2}\vec{k_{\perp}} \over \pi k_{\perp}^{2}}
 {dG(\bar{x}, k_{\perp}^{2}) \over d\log k_{\perp}^{2}}
d\sigma_{hard}(\vec{p}_{J},\vec{k}_{\perp},....)\,,
\label{eq:2}
\end{equation}
where $d\sigma_{hard}(\vec{p}_{J},\vec{k}_{\perp},....)$ must include the
transfer of the intrinsic $k_{\perp}$ to the produced partons/jets.
For instance, in the $\gamma^{*}g \to q\bar{q}$ fusion the two jets 
will have the transverse 
momenta $\vec{p}_{J1}$ and $\vec{p}_{J2}=\vec{p}_{J1}+\vec{k_{\perp}}$
and, to a  crude approximation, the effect of $\vec{k}_{\perp}$ is a
smearing 
\begin{eqnarray}
{d\sigma_{J}(\vec{p}_{J})\over d^{2}\vec{p}_{J}} =
{1\over 2}\left\{
\left.{d\sigma_{J}(\vec{p}_{J})\over d^{2}\vec{p}_{J}} \right|_{unsmeared}
+ {1\over G(\bar{x}, p_{J}^{2}) }
\int {d^{2}{\vec{k_{\perp}}} \over \pi k_{\perp}^{2}}
\left. {dG(\bar{x}, k_{\perp}^{2}) \over d\log k_{\perp}^{2}}
{d\sigma_{J}(\vec{p}_{J}-\vec{k}_{\perp})\over d^{2}\vec{k}_{J}}
\right|_{unsmeared} \right\}\, ,
\label{eq:3}
\end{eqnarray}
where the unsmeared term is similar to that of the 
$Q^{2}$-factorization. Evidently, the $k_{\perp}$-smearing over the broad
$k_{\perp}$-distribution strongly enhances the observed jet cross section.

Surprisingly enough, even 20  years after FKL the idea of a
consistent use of the FKL unintegrated structure functions did not
permeate yet the pQCD parton model phenomenology. On the experimental side,
it has been well established that neither various $p_{\perp}$ distributions 
nor azimuthal correlations between high-$p_{\perp}$ particles can be
understood quantitatively in the collinear parton approximation. The
LO collinear approximation fails badly, NLO calculations fare somewhat
better because they introduce a semblance of intrinsic $k_{\perp}$ via 
initial state radiation. Still, this initial state radiation is {\sl very much 
insufficient} and must be complemented by an {\sl artificial and very large } 
intrinsic $k_{\perp}$. An excellent illustration of this point is
provided by the recent FNAL E706 data on direct photons and neutral mesons
at $3.5 < p_{\perp} < 11$ GeV in fixed target 515 GeV $\pi^{-}Be$ and 530 \& 800 GeV
$pBe$ collisions \cite{E706}. E706 finds that the standard NLO predictions 
fall short of the observed cross section by a factor 3-5.
Because in this case one can not fiddle with the gluon density, 
the theory can be brought to agreement with 
experiment only at the expense of endowing colliding partons with a very large intrinsic 
transverse momentum, $\langle k_{\perp} \rangle  \approx 1.2-1.5$ GeV/c.
Equally large supplemental $\langle k\rangle_{\perp}\sim 1.5$ GeV/c is 
necessary for a good description of the E706 data on azimuthal correlation 
of pairs of high-$p_{\perp}$ pions. All the above shows that neglecting the well defined
dynamical $k_{\perp}$ of partons is not warranted.\\

\noindent
{\bf 3.~~ Dynamical intrinsic $k_{\perp}$ in diffractive DIS}\\

The driving QCD subprocesses of diffractive DIS are elastic 
$\gamma^{*}p\to Xp'$ and/or proton-dissociative $\gamma^{*}p \rightarrow XY$ 
excitation of $X=q\bar{q}$ and $X=q\bar{q}g$  Fock states of the photon. The 
complete and physically very transparent QCD description of these processes in the 
color dipole representation has been given in  \cite{GNZ95,NZ94,NZ92}.
Of course, the Fourier transform allows to cast the same results in 
 momentum space too \cite{GNZcharm,GNZlong,NZsplit,Wusthoff}.
In the both languages, the dynamics of diffraction is controlled by the
gluon structure function of the proton $G(x,q^{2})$. 

The jet transverse momentum is measured with respect to 
the $\gamma^{*}\Pom$ collision axis. 
Typically neither
the recoil proton $p'$ nor proton-dissociative state $Y$ are observed and
momentum analyzed. However, if one parameterizes the dependence on 
the $(\gamma^{*},X)$ momentum transfer $\Delta$ as
$
{d\sigma_{D}/ d\Delta^{2}} \propto \exp(-B_{D}\Delta^{2})\, ,
$
then for elastic case the diffraction slope $B_{D}\sim 
6$ GeV$^{-2}$ \cite{NPZslope}, whereas in 
the proton-dissociative case $B_{D} \sim $2 GeV$^{-2}$ as was argued in 
\cite{HNSSZ}. Therefore, typical
$\Delta$ are small and uncertainties with the $\gamma^{*}\Pom$
axis can be neglected.   

Excitation of the $q\bar{q}$ states gives rise to back-to-back jets aligned
predominantly along the $\gamma^{*}\Pom$ collision and dominates diffractive
DIS at $\beta \gsim 0.1$ \cite{GNZ95,NZ92}. The $p_{J}$ distribution of 
these jets was calculated in \cite{NZ92,NZsplit} and there are two distinct
regimes. At $\beta \sim 0.5$ the transverse momentum of
the two gluons which form the exchanged pomeron does not contribute to 
$p_{J}$ of the jet and one finds
\begin{equation}
{d\sigma_{D} \over  dM^{2} d\Delta^{2}dp_{J}^{2}} \propto
{1\over p_{J}^{4}}G^{2}({1\over 2}x_{\Pom},p_{J}^{2})\, .
\end{equation}
The situation changes dramatically at $\beta \ll 1$. In this case $p_{J}$
comes entirely from the transverse momentum of exchanged gluons and
\begin{equation}
{d\sigma_{D} \over  dM^{2} d\Delta^{2}dp_{J}^{2}} \propto
{1\over p_{J}^{4}}\left[{dG^{2}({1\over 2}x_{\Pom},p_{J}^{2})\over d\log p_{J}^{2}}
\right]^{2}\, .
\end{equation}
In \cite{NZsplit} we dubbed this process `splitting of pomerons into two jets'.
For real photons the splitting of pomerons is a dominant mechanism of 
diffractive back-to-back high-$p_{J}$ jet production and this is yet unexplored
process is a unique direct probe of the FKL unintegrated gluon structure function
of the proton.
 
To standard leading log$Q^{2}$ approximation 
diffractive excitation of the $q\bar{q}g$ can be viewed as
a photon-gluon fusion $\gamma^{*}g \to q\bar{q}$ where the gluon comes from
the valence gluon-gluon state of the Pomeron. The crucial finding behind
this interpretation is that the $q\bar{q}$ separation in the impact 
parameter space is much smaller than the $qg$ ($\bar{q}g$) separation
\cite{NZ94}.
In the momentum space that corresponds to the conventional DGLAP ordering
of the transverse momenta 
$
k_{g}^{2} \ll k_{q}^{2},k_{\bar{q}}^{2} \ll Q^{2}$
which is the basis of the proof \cite{NZ94} of the DGLAP evolution of the
pomeron structure function at small $\beta$.
 Furthermore in \cite{NZ94} we have derived the spatial wave function of 
the gluon-gluon state of the Pomeron which allows to quantify the distribution
of the intrinsic transverse momentum of gluons $k_{g}$ in the pomeron, the gross
features of which are given by 
\begin{equation}
{dg_{\Pom}(z_{g},k_{g}^{2}) \over dk_{g}^{2}}= 
|\Psi(z,\vec{k}_{g},x_{\Pom})|^{2})
\sim {(1-z_{g})^{\gamma} \over z_{g}}\cdot {1 \over (k_{g}^{2} +\mu_{G}^{2})^{2}}
\cdot G^{2}({1\over 2}x_{\Pom},k_{g}^{2}+\mu_{g}^{2})\, ,
\label{eq:5}
\end{equation}
where $z_{g}$ is a fraction of the pomeron's momentum carried by the
gluon. Here $\mu_{G}$ is the fundamental infrared cutoff related to
the propagation radius $R_{G}=\mu_{G}^{-1}$ of perturbative gluons. 
The lattice QCD and other model studies suggest $R_{G} \sim $0.2-0.3 fm and
the rather large $\mu_{G}=R_{c}^{-1}\sim$ 0.7-1 GeV. 
This infrared parameter $\mu_{G}$ defines the onset of pQCD: only gluons with 
$k_{g} >> \mu_{G}^{2}$ can be treated as perturbative ones, the large value 
of $\mu_{G}$ explains why DGLAP de-evolutions to small $Q^{2}$ run into trouble
at $Q^{2} \sim $ 1-1.5 GeV$^{2}$. The large value of $\mu_{G}$ entails large
intrinsic $k_{\perp}$ of gluons in the pomeron, and the tail of the $k_{g}$
distribution is very strongly enhanced by the scaling violations, especially
at very small $x_{\Pom}$.

Notice that the gluon distribution in pomerons (\ref{eq:5}) 
depends on $x_{\Pom}$, which is a manifestation of the well understood
but often ignored lack of Regge factorization \cite{GNZ95,GNZcharm}, 
because of which the pomeron
can not be treated as a hadronic state.

For some time H1 keeps advocating \cite{H1} flat, $\gamma\sim 0$, or even strongly
peaked at $z_{g}\sim 1$ pomeron glue. None are 
possible in the QCD mechanism of diffraction which suggests 
\cite{GNZ95,NZ94,Wusthoff}
rather soft valence glue with $\gamma \sim $2-3.

The impact of the long tail of the $k_{g}$ distribution on the jet structure
of final states 
is obvious. One looks for jets in events with large transverse energy $E_T$ and 
intrinsic $k_{g}$ enhances $E_{T}$ by $\sim 2k_{g}$. To the lowest order in
pQCD, the typical final $q\bar{q}g$ state would consist of three jets: the 
Pomeron remnant jet with $\vec{p}_{J1}=-\vec{k}_{g}$ with respect to the 
$\gamma^{*}\Pom$ collision axis and the transverse momentum distribution 
given by Eq. (\ref{eq:5}) and the two 
jets from photon-gluon fusion with the transverse momenta $-\vec{p}_{J2}$ and
$\vec{p}_{J3}=\vec{p}_{J2}+\vec{k}_{g}$.

 The $k_{g}$ distribution 
(\ref{eq:5}) gives the unintegrated gluon density in the pomeron, which 
must be used in evaluations of the smearing effect in the photon-gluon 
fusion following (\ref{eq:3}). This smearing will enhance substantially the 
predicted jet cross section. Inverting the problem, if the $k_{\perp}$-smearing
were included, then one would reproduce the same observed jet cross section
with smaller gluon density in the pomeron than in the standard $Q^{2}$-factorization
without $k_{\perp}$-smearing. Diffraction excitation of higher $q\bar{q}gg...$ 
Fock states of the photon 
complemented by virtual radiative corrections to excitation of lower Fock states
amounts to the conventional QCD evolution of the
glue in the pomeron starting with 
the input glue (\ref{eq:5}). As usual, this evolution will steepen the small-$z_{g}$
behavior and still further broaden the gluon $k_{\perp}$-distribution. 

The most direct experimental confirmation of the above $k_{\perp}$-smearing would be
azimuthal decorrelations of jets `2' and `3' in the photon hemisphere.  Let 
$k_{out}$ be the component of $\vec{p}_{J3}$ out of the plane formed by the
momentum $\vec{p}_{J2}$ and the $\gamma^{*}\Pom$ collision axis. Then {\bf we predict}
the $k_{out}$ distribution of the form
$$
{dN\over dk_{out}} \propto {1 \over (k_{out}^{2}+\mu_{G}^{2})^{3\over 2}}
\cdot G^{2}({1\over 2}x_{\Pom},k_{out}^{2}+\mu_{g}^{2})\, ,
$$
for $k_{out} \lsim p_{J}$, which can be tested experimentally. Incorporation of the
intrinsic $k_{g}$ distribution (\ref{eq:5}) into RAPGAP and other generators must not
be a problem and, hopefully, will result in a good description of the diffractive 
jet data with less glue in the pomeron.

{\bf Acknowledgments:} Thanks are due 
to R.Roosen for invitation to present the above ideas at the Workshop,
to H. Jung for explanations on RAPGAP and to extremely helpful 
suggestions on the organization of the manuscript and to M.Diehl for
useful comments.

\end{document}